\documentclass[11pt,twoside]{article}

\usepackage{asp2004}
\usepackage{epsf}
\usepackage{psfig}
\usepackage{lscape}

\begin{document}
\title{Radio astronomical probes of
cosmic reionization and the first luminous sources:
probing the 'twilight zone'}   
\author{C.L. Carilli}  
\affil{NRAO PO Box O Socorro NM USA}

\begin{abstract} 

The epoch of reionization (EoR) corresponds to a 'cosmic phase
transition', when the neutral intergalactic medium (IGM) becomes
ionized by the first stars and/or AGN. While the discoveries of
Gunn-Peterson (GP) absorption troughs in the spectra of the highest
redshift QSOs, and large scale polarization of the CMB, have set the
first hard constraints on the EoR, the redshift and process of
reionization, and the nature of the first luminous objects, remain two
of the paramount questions in cosmic structure formation. Moreover,
the GP effect is such that observations of objects during this epoch
will be difficult at wavelengths shorter than about 1 micron.  Hence,
cosmic reionization, and the formation of the first luminous objects,
occurs in a 'twilight zone', observable only at radio through near-IR
wavelengths.  In this talk I explore studies of the EoR at meter
through submillimeter wavelengths. I present recent observations of
the dust, molecular gas, and star formation activity in the host
galaxies of the highest redshift QSOs. These results have interesting
implications on the timescale for metal and dust enrichment, on the
possibility of co-eval formation of SMBHs and galaxies, and on the
process of reionization. I then discuss future capabilities of low
frequency radio astronomy to study the neutral IGM via the HI 21cm
line, including imaging and power spectral analyses of large scale
structure in emission, and absorption studies toward the first radio
loud sources. I conclude with a summary of the VLA-VHF system to study
cosmic Stromgren spheres associated with the highest redshift SDSS
QSOs in the HI 21cm line at 190 MHz.

\end{abstract}

\section{Introduction}  

Cosmic reionization corresponds to the epoch when the neutral
intergalactic medium (IGM) is reionized by the first luminous sources
(stars or accreting black holes), after having coasted in a neutral
state from the time of recombination. The existence of a neutral IGM
prior to the formation of the first luminous objects is perhaps the
last major phase of cosmic evolution that remains to be
verified. Also, discovery of the epoch of reionization (EoR) provides
a fundamental benchmark in cosmic structure formation, indicating the
formation of the first luminous objects (Loeb \& Barkana 2000).

The last few years have seen the first direct observational
constraints on cosmic reionization.  First, discovery of Gunn-Peterson
absorption troughs in the spectra of the highest redshift QSOs,
corresponding to Ly-$\alpha$ absorption by the neutral IGM, suggest an
increase in the cosmic neutral fraction from f(HI) $\le 10^{-4}$ at $z
< 5.5$, to f(HI)$ > 10^{-3}$ at $z \ge 6$ (Fan et al. 2003; White et
al.  2003).  Relatively late reionization is supported by study of the
thermal state of the IGM at $z \sim 5$ through QSO absorption lines
(Hui \& Haiman 2003). On the other hand, the large scale polarization
signal of the CMB detected by WMAP, resulting from Thomson scattering
during the EoR, suggests a significant ionization fraction for the IGM
(f(HI)$> 0.5$) out to high redshift ($z \sim 17$; Kogut et al. 2003). It
has also been argued that the discovery of Ly$\alpha$ emitting
galaxies at $z > 6$ implies a significantly ionized IGM (Stern et
al. 2005; Malhotra \& Rhoades 2004), although source clustering may
alleviate this requirement (Furlanetto et al. 2004; Haiman 2002).

Overall, these first observational constraints on the EoR suggest that
reionization may be a complex process extending from $z \sim 17$ down
to $z \sim 6$. As an example of models being considered, perhaps the
simplest is the 'double reionization' model of Cen (2002), in which
the universe is reionized at $z \sim 17$ by massive pop III stars
associated with the first non-linear objects exceeding the cosmic
Jean's mass (mini-halos $\sim 10^{6-7}$ M$_\odot$). This process is
self-limiting, since the Pop III stars will both disrupt the host
galaxies when they go supernovae (Abel et al. 2002), as well as warm
the IGM, thereby increase the Jean's mass. Once the sources are turned
off, the IGM density at $z>8$ is high enough that the universe has
time to recombine (although cf. Furlanetto \& Loeb 2004).  A second
reionization occurs at $z \sim 6$ to 7 due to normal star formation in
galaxies with masses $\ge 10^8$ M$_\odot$.

These first observational constraints on the EoR represent major
advances on our understanding of cosmic structure formation.  However,
each has its limitations.  Considering the GP effect, spectra in the
0.8 to 1 $\mu$m range are severely contaminated by night sky OH lines,
and the conclusions depend on an extrapolation of QSO UV spectra below
Ly$\alpha$. Most importantly, the universe becomes opaque at fairly
low neutral fractions (f(HI) $\sim 10^{-3}$), and hence the technique
loses diagnostic capability. Indeed, Songaila (2004) has recently
challenge the turn-up in the opacity at $z \sim 6$, suggesting a more
gradual rise.  The WMAP large scale polarization measurement is
effectively an integrated measure of the universal Thomson optical
depth, and hence can be fit by myriad physical models for reionization
(Gnedin 2005). Also, there are significant uncertainties involved in
subtracting the (possibly) polarized foregrounds (eg. the Galaxy) on
the large scale being considered (10's of degrees).

The coming years should see considerable effort in improving the
observational constraints on the EoR, and in reconciling these
measurements with physical models for cosmic structure formation.  In
this talk I summarize radio astronomical constraints on cosmic
reionization, excluding the CMB (see also Carilli et al. 2004a). These
constraints include: (i) observations at cm through (sub)mm
wavelengths of emission from the dust, gas, and star formation, in the
highest redshift objects, and (ii) study of the neutral IGM through
observations of the HI 21cm line with future large area low frequency
radio telescopes.

\section{Radio studies of objects within the 'twilight zone'}

The onset of the GP effect at $z\sim 6$ implies that the universe
becomes opaque at (observed) optical wavelengths. Hence, cosmic
reionization, and the formation of the first luminous objects, occurs
in a 'twilight zone', observable only at radio through near-IR
wavelengths ($\lambda > 0.9\mu$m), and in the hard Xrays.

What is the 'magic' of (sub)mm observations of the high redshift
universe?  Figure 1a shows the radio through far-IR (FIR) spectrum of
the starburst galaxy Arp 220.  The cm emission corresponds to
non-thermal (synchrotron) emission from relativistic electrons
accelerated in SNR shocks, while the mm-FIR emission corresponds to
thermal emission from warm dust heated by the interstellar radiation
field.  The fact that both emission mechanisms (non-thermal and
thermal) are driven by (massive) star formation leads to the well
known linear correlation between the radio and FIR luminosities of
star forming galaxies (Yun et al. 2001). The (sub)mm spectral range
also includes myriad spectral lines from rotational transitions of
common interstellar molecules, as well as low excitation electronic
fine structure lines from common elements.

Figure 1b shows the flux density of Arp 220 as a function of redshift
at observing frequencies of 250 and 350 GHz.  The large 'inverse-K'
correction due to the rapid rise on the Rayleigh-Jeans side of the
gray body spectrum leads to a roughly constant flux density from $z
\sim 0.5$ to 8 for a star forming galaxy. Hence, (sub)mm observations
provide a uniquely {\sl distance independent} method for studying
objects throughout the universe.

\begin{figure}
\psfig{figure=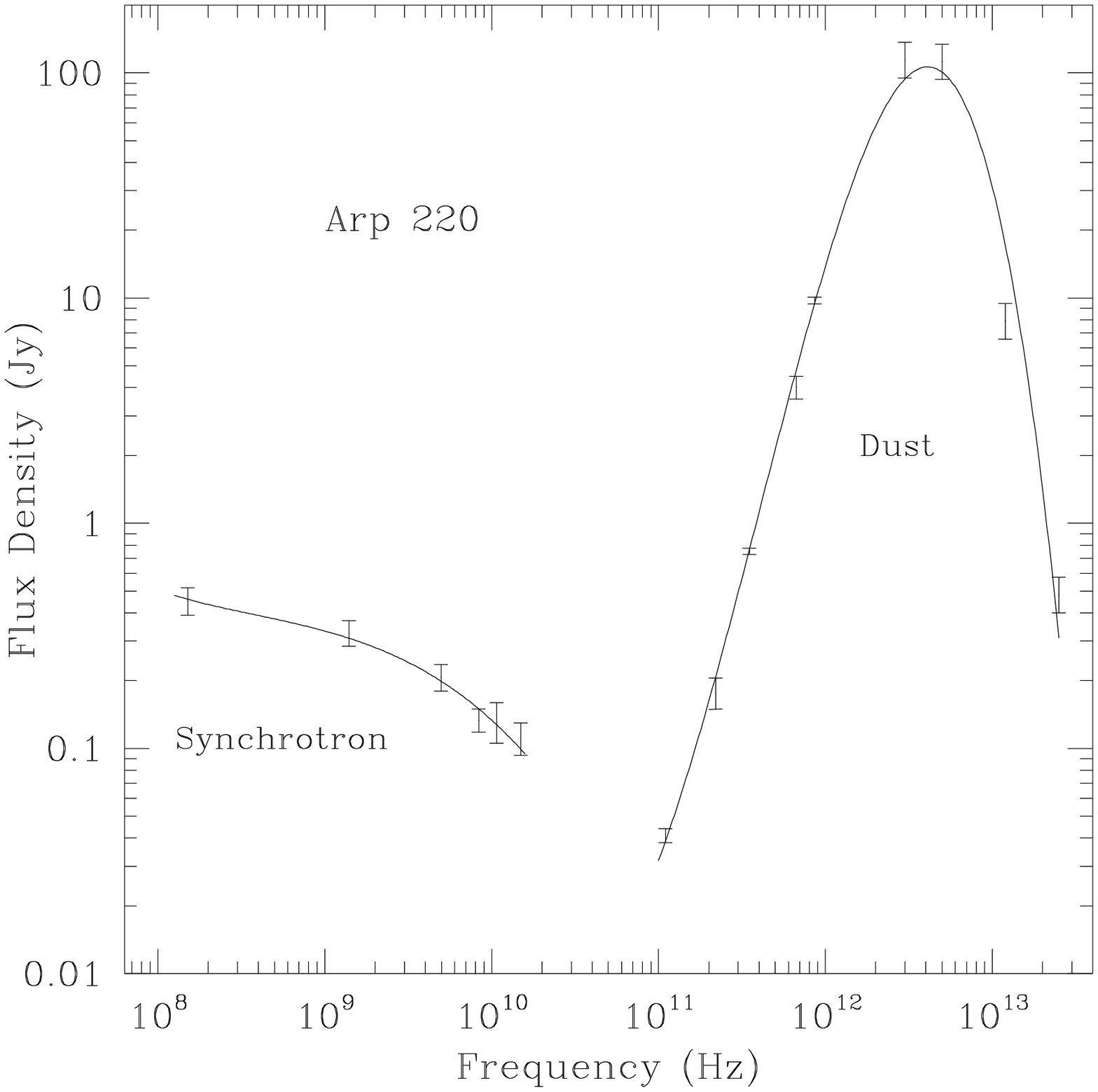,width=2.7in}
\vspace*{-2.7in}
\hskip 2.7in
\psfig{figure=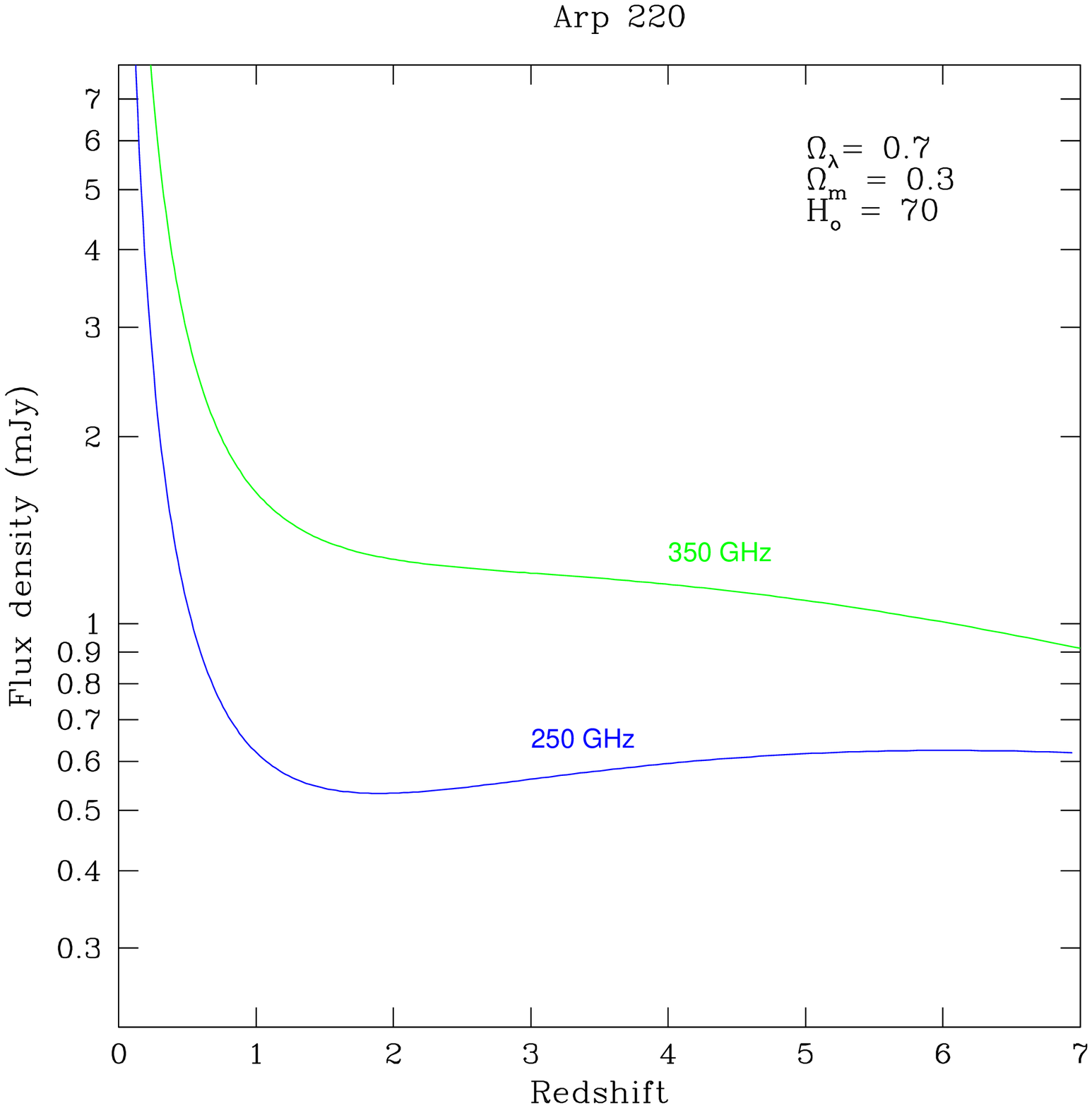,width=2.7in}
\caption{{\bf Left:} The spectral energy distribution of 
Arp 220 from the radio through IR. {\bf Right:} The expected
flux density at fixed observing frequencies (250 and 350 GHz)
for Arp 220 as a function of redshift. }
\end{figure}

\subsection{High redshift QSOs}

QSO host galaxies have long been targets for studies of high $z$
dust and molecular line emission, because of the fact that
optical redshifts are easily measured, as compared to  submm
galaxy samples, for which obtaining optical redshifts has been
notoriously difficult (Blain et al. 2002). 

The past decade has seen a revolution in our understanding of high
redshift QSOs in three ways. First, surveys such as the SDSS and DPOSS
have revealed large samples of QSOs to the highest redshifts, with
close to 1000 beyond $z = 4$, and 8 beyond $z = 6$ (Fan et
al. 2003; Djorgovski et al. 2001).  Due to sensitivity limits, these
QSOs are all extreme luminosity sources, with M$_B < -26$, and
bolometric luminosities (dominated by the 'big blue bump') $> 10^{14}$
L$_\odot$.  Eddington limited accretion implies black hole masses $>
10^9$ M$_\odot$.

Second, study of QSO host galaxies at lower redshifts has revealed a
correlation between the mass of the black hole and the velocity
dispersion of the host spheroidal galaxy (Ferrarese \& Merritt 2000;
Gebhardt et al. 2000).  This $\rm M_{BH} - \sigma$ relation implies
that most (all?) spheroidal galaxies contain supermassive black holes,
and that the hole mass correlates with the mass of the host galaxy:
$\rm M_{BH} \sim 0.002 M_{bulge}$, leading Gebhard et al. (2000) to
hypothesize a 'causal relationship' between the formation of
supermassive black holes and spheroidal galaxies.  This correlation
also implies that the highest redshift QSOs are associated with
massive host galaxies ($\sim 10^{12}$ M$_\odot$), if the relationship
holds to $z \sim 6$.

And third is the discovery that roughly 1/3 of optically selected QSOs
are detected at (sub)mm wavelengths in surveys with mJy sensitivity
(Beelen 2005; Omont et al. 2003; Carilli et al. 2004c), and that this
fraction remains constant out to the highest redshifts (Bertoldi et
al. 2004).  Beelen shows that the FIR luminosity correlates weakly
with the bolometric luminosity of the QSO. Studies of the
radio-through-FIR SEDs of these sources show that most follow the
radio-FIR correlation established by low redshift star forming
galaxies (Petric et al. 2004; Carilli et al. 2004b), while VLBI
observations of the radio continuum emission show brightness
temperatures $\le 10^4$ K, again consistent with starburst galaxies
(Momjian et al 2004; Beelen et al. 2004).  On the other hand, the FIR
luminosity is typically only 10$\%$ of the bolometric luminosity, such
that dust heating by the AGN cannot be precluded (Andreani et al.
2003).

\begin{figure}[htb]
\psfig{figure=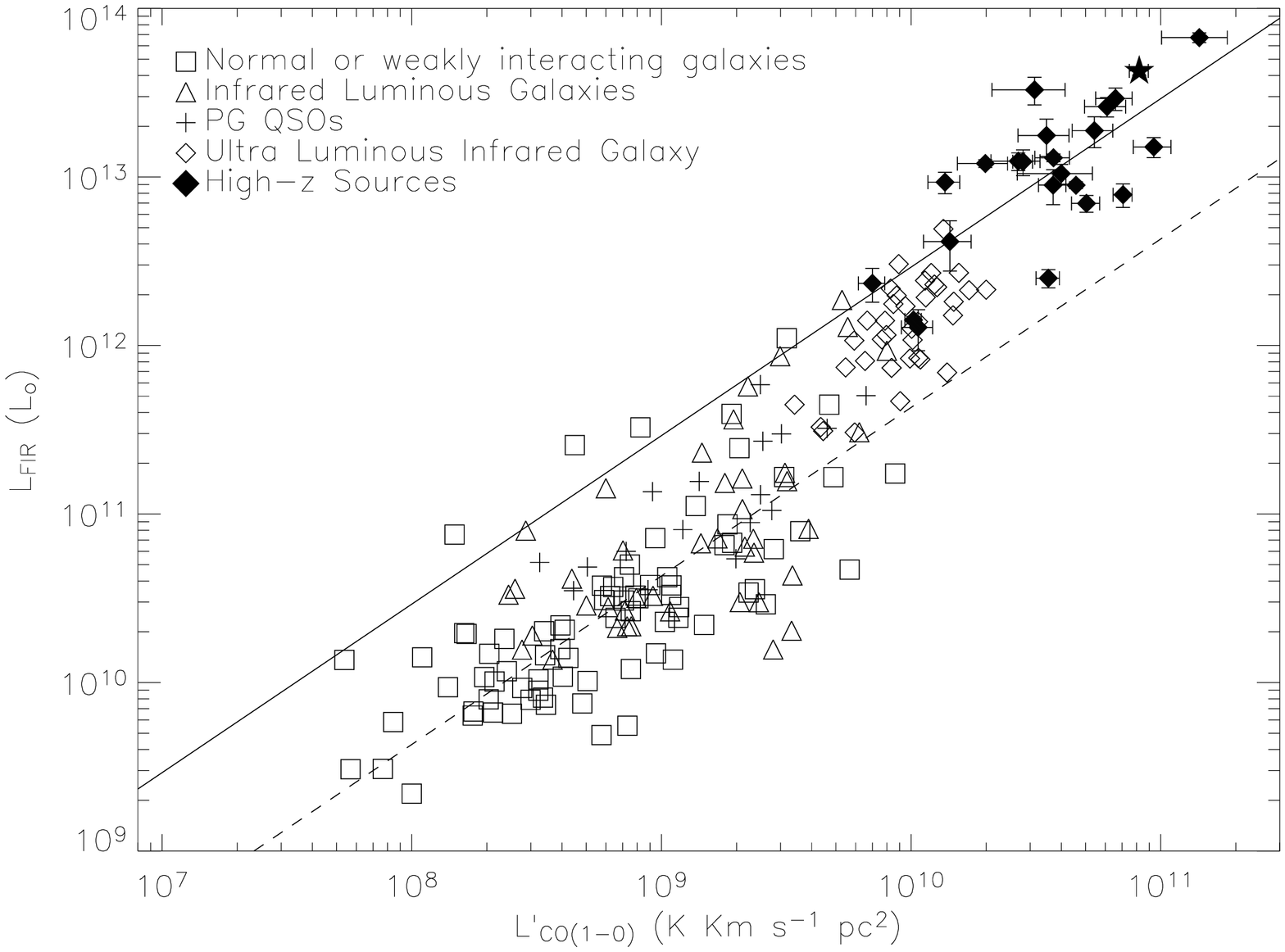,angle=90,width=2.7in}
\vspace*{-2.3in}
\hskip 2.5in
\psfig{figure=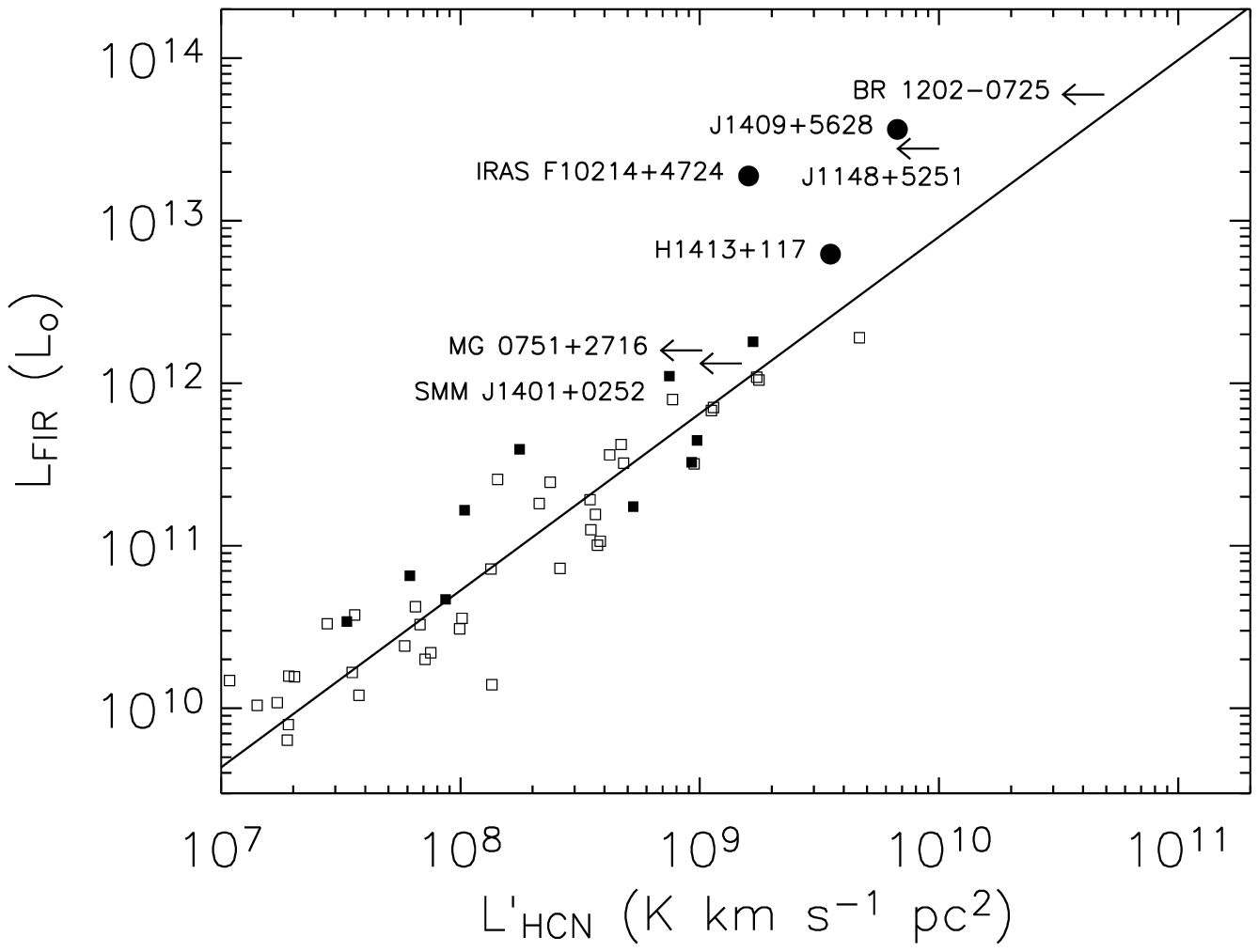,width=3.4in}
\caption{{\bf Left:} The correlation between FIR and CO luminosity
for low $z$ star forming galaxies (open) and high $z$
galaxies and AGN (solid). The solid and dash 
lines are powerlaws of index 1. {\bf Right:} The same, but for 
HCN luminosity (Beelen 2005; Carilli et al. 2005).}
\end{figure}

Figure 2a shows the relationship between FIR luminosity and CO
luminosity for a large sample of low $z$ galaxies, and for the high
redshift sample, comprised mostly of QSO host galaxies.  It is well
known that low $z$ galaxies follow a tight, but non-linear,
correlation between L$_{\rm FIR}$ and L$'_{\rm CO}$, consistent with a
powerlaw of index = 1.7 (Gao \& Solomon 2004). The FIR luminosity can
be interpreted as a measure of star formation rate: $\rm SFR \sim
4\times 10^{-10}~ L_{\rm FIR}~ M_\odot~ year^{-1}$, with L$_{\rm FIR}$
in L$_\odot$. The CO luminosity has been used as a empirical
diagnostic of molecular (H$_2$) gas mass: $\rm M(H_2) = X~ L'(CO)$,
with X in M$_\odot$ (K km s$^{-1}$ pc$^2$)$^{-1}$. For Galactic GMCs
the empirical value of X = 4.6 (Dame et al. 1987), while for starburst
galaxies X = 0.8 (Downes \& Solomon 1998).  The non-linear
relationship between L$_{\rm FIR}$ and L$'(\rm CO)$ has been
interpreted as an increasing 'star formation efficiency' (= SFR/gas
mass) with increasing gas mass (Gao \& Solomon 2004).  The fact that
the high redshift sources follow the same (non-linear) relationship
between L$_{\rm FIR}$ and L$'(\rm CO)$ suggests a similar dust heating
mechanism in each, and in the low $z$ galaxies the dust heating
mechanism is known to be star formation.

The CO molecule has a low dipole moment, and hence a fairly low
critical density for excitation ($\sim 10^3$ cm$^{-3}$).  Hence CO
emission traces both diffuse and dense molecule clouds.  The next
strongest molecular emission line from starburst
galaxies is that from HCN.  Figure 2b shows the relationship between
L$_{\rm FIR}$ and L$'(\rm HCN)$ (Gao \& Solomon 2004).  HCN is a high
dipole moment molecule, with a critical density for excitation $\sim
10^5$ cm$^{-3}$. Hence HCN traces only dense gas associated with
actively star forming molecular clouds.  Again, a correlation is seen
for low $z$ star forming galaxies, but in this case the relationship
is linear (power law index = 1), consistent with the idea that HCN
traces only the densest gas directly associated with star forming
clouds (Gao \& Solomon 2004).  The different powerlaw indices for the
correlations of CO and HCN with FIR luminosity also imply that the
HCN/CO luminosity ratio increases with FIR luminosity, from a few$\%$
at FIR $\sim 10^{10}$ L$_\odot$ up to 10 to 20$\%$ for $10^{12}$
L$_\odot$.

Figure 2b shows that the high $z$ QSO host galaxies roughly follow
this linear trend between HCN and FIR luminosity, consistent with star
forming clouds (Carilli et al. 2005).  However, we should point out
that many of the measurements for the QSO hosts are upper limits in
HCN, and that in all cases the points (or limits) fall on the low-side
of the curve with respect to L$\rm (FIR)$. This could be used to argue
for a contribution to dust heating by the AGN.

Searches have also been made for low order atomic fine structure lines
from QSO host galaxies, known to be the dominant cooling lines in the
ISM.  Thus far the results have been disappointing, with limits to the
line luminosities of eg. the [CII] 158 $\mu$m line falling well below
the expected value extrapolated from the FIR luminosity (van der Werf
1991). However, a similar under-luminosity for these atomic submm lines
has been seen in low $z$ ultra-luminous IR galaxies (Dale et
al. 2004). This under-luminosity has been interpreted as resulting
from a decreased efficient for photoelectric heating by charged dust
grains in starburst environments.

Recently Pety et al. (2005) have detected the [CI] 609$\mu$m line at
$z = 4.12$ from the gravitationally lensed FIR-luminous QSO PSS
2322+1944. They conclude that the ISM of the QSO host galaxy
must have roughly solar metalicity. 

Overall, cm and mm observations of high $z$ QSO host galaxies in the
dust continuum emission, molecular lines, fine structure lines, and
non-thermal radio continuum, are consistent with starbursts in the
host galaxies concurrent with the AGN activity. If the dust heating is
by star formation, the star formation rates are of order $10^3$
M$_\odot$ year$^{-1}$, such that a large spheroidal galaxy could form
over a dynamical timescale of 10$^8$ years. 

\subsection{J1148+5251: the highest redshift QSO}

The highest redshift QSO known is SDSS J1148+5251 at $z = 6.42$ (Fan
et al. 2003). The bolometric luminosity of this source is $\sim
10^{14}$ L$_\odot$, with a supermassive black hole of a
few$\times10^9$ M$_\odot$ (Willott et al. 2004).  J1148+5251 is also
one of the 30$\%$ of the high $z$ QSOs that is also FIR luminous,
being detected at 250 GHz with MAMBO with an implied FIR luminosity
$\sim 1.3\times10^{13}$ L$_\odot$, and a dust mass of $8\times10^8$
M$_\odot$ (Bertoldi et al.  2003).

\begin{figure}[htb]
\psfig{figure=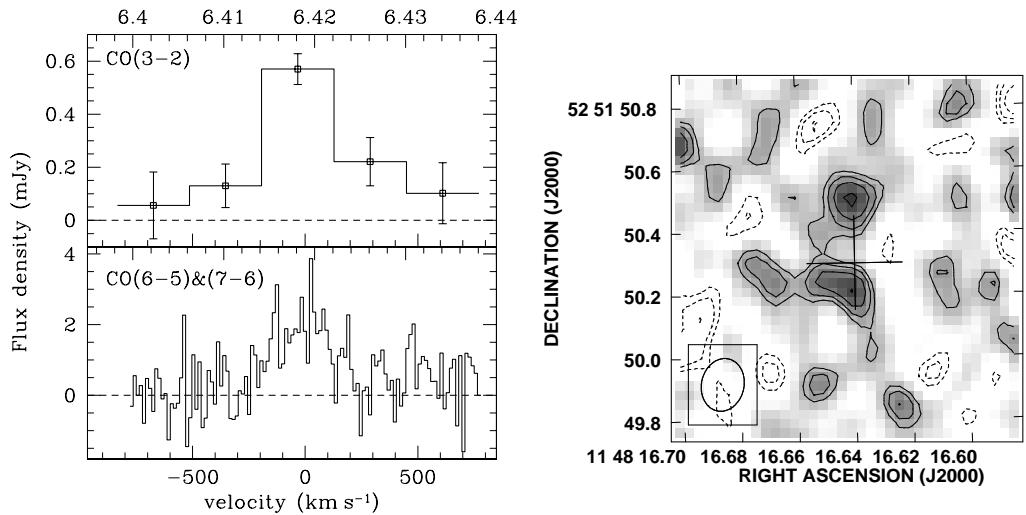,width=2.7in}
\vspace*{-2.4in}
\hskip 2.7in
\psfig{figure=f3.eps,width=2.7in}
\caption{{\bf Left:} The CO 3-2 and 6-5 emission from J1148+5251 at $z= 6.42$.
{\bf Right:} The  image of the CO 3-2 emission made using
the VLA at 0.15$''$ resolution (Walter et al. 2003, 2004; Bertoldi et al.
2003). The cross shows the position of the optical QSO (White
et al. 2005).}
\end{figure}

CO emission has been detected from the host galaxy of J1148+5251 with
both the VLA (3-2 line) and PdBI (6-5, 7-6 lines; Walter et al. 2003;
Bertoldi et al. 2003; Fig. 3a). The implied molecular gas mass is
$2\times 10^{10}$ M$_\odot$. The gas-to-dust ratio is $\sim 30$,
similar to that seen in active star forming galaxies The CO excitation
follows roughly constant brightness temperature up to 6-5, indicating
dense $\sim 10^5$ cm$^{-3}$, warm ($\sim 100$ K) gas, again similar to
that seen in starburst galaxy nuclei. The line profile is roughly
Gaussian, with a FWHM = 305 km s$^{-1}$ (Bertoldi et al. 2003).

High resolution imaging of the CO emission with the VLA reveal
molecular gas extended over about 1$''$, or 5.5 kpc, with about half
the emission coming from two compact components separated by $0.3''$
(=1.7 kpc; Walter et al. 2004; Fig. 3b).  Walter et al. consider the
possibility of a double nucleus, as could occur in a recent galaxy
merger. They also calculate the dynamical mass from the observed
line width and spatial distribution, and find a dynamical mass
within 2.5 kpc radius of $5\times 10^{10}$ M$_\odot$. The intrinsic
CO brightness temperature is $\sim 20$K for each component, 
which is typical for CO emission from starburst nuclei (Downes
\& Solomon 1998). 

The dynamical mass and gas mass of J1148+5251 are comparable,
suggesting that the gravitational mass in the inner few kpc of the
host galaxy is baryon dominate, as is also seen for nearby elliptical
galaxies, although in the case of 1148+5251 the mass is in gas, while
in low $z$ galaxies the mass is in stars.  From the $\rm M_{BH} -
\sigma$ relation, Walter et al. derive a bulge mass of $1.5\times
10^{12}$ for J1148+5251. This value is much larger than the dynamical
mass, leading Walter et al. to suggest that the $\rm M_{BH} -
\sigma$ relation breaks down at the highest redshifts.  The indication
is that the SBMH forms prior to the host spheroidal galaxy (Wyithe \&
Padmanaban 2005).

J1148+5251 has also been detected at 1.4 GHz with the VLA,
corresponding to non-thermal (synchrotron) emission. The
radio-through-IR spectrum is shown in Fig. 4. The spectrum is
consistent with that expected for an active star forming galaxy,
ie. the host galaxy emission follows the radio-FIR correlation for
star forming galaxies.  Overall, the physical conditions measured for
J1148+5251, from the dust, molecular gas, and non-thermal continuum
emission, are consistent with a massive starburst in the host galaxy,
coeval with the AGN activity, with an implied star formation rate
$\sim 10^3$ M$_\odot$ year$^{-1}$, perhaps in a merging galaxy
system. 

\begin{figure}[htb]
\psfig{figure=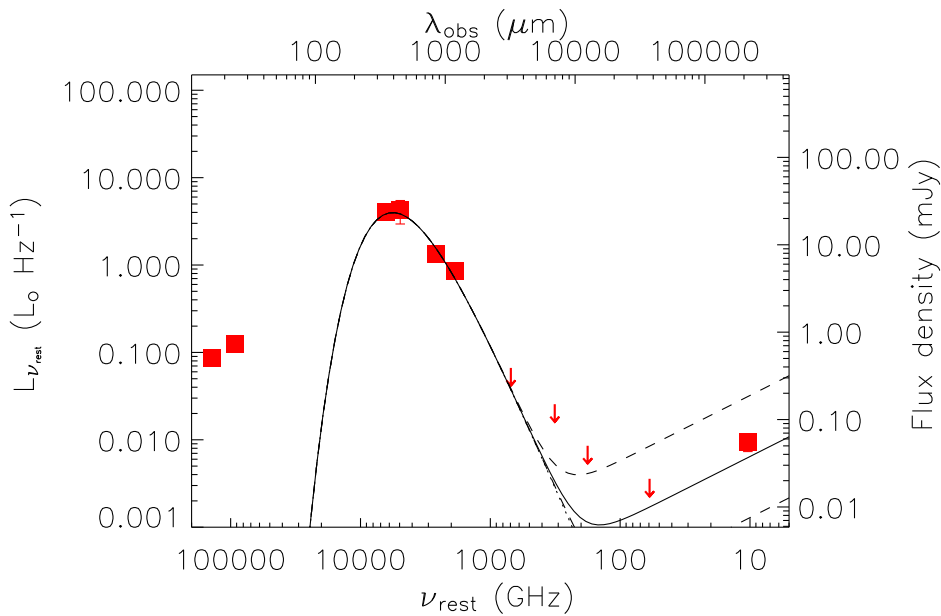,width=5in}
\caption{The near-IR through radio 
spectral energy distribution (rest frame) for J1148+5251
at $z=6.42$ (Beelen et al. 2005).  The solid curve shows a typical SED
of a star forming galaxy with a dust temperature of 50 K, and the
dash lines show the range that corresponds to the radio-FIR correlation
for star forming galaxies (Yun et al. 2001). 
}
\end{figure}

The age of the universe at $z = 6.4$ is only $0.87\times 10^9$ years.
This short timescale precludes the standard ISM dust formation
mechanism via winds from evolved lower mass stars (ie. AGB stars),
which requires $> 1.4 \times 10^9$ years. A number of authors have
considered the possibility of dust formation within supernovae or
supernovae remnants, which could operate on a much shorter timescale
(Dunne et al. 2003; Maiolino et al. 2004). 

\section{Cosmic Stromgren spheres and the state
of the IGM at $z > 6$}

The CO observations of J1148+5251 provide a very accurate redshift for
the host galaxy: $z = 6.419 \pm 0.001$.  For comparison, redshifts
based on high ionization broad lines provide very inaccurate host
galaxy redshifts, with offsets up to 2000 km s$^{-1}$ (Richards et
al. 2002). Low ionization broad lines, such as MgII, are thought to
provide more accurate host galaxy redshifts (within a few hundred km
s$^{-1}$), although clearly not as accurate as the CO line.

Why is an accurate host galaxy redshift important?  White et
al. (2003) point out that, if the IGM was neutral (or even 0.1$\%$
neutral) right up to the host galaxy redshift, then there should be no
emission seen blue-ward of Ly$\alpha$.  And yet, sensitive spectra of
$z>6$ QSOs show emission extending below the nominal host galaxy
redshift before the Gunn-Peterson absorption trough sets-in. This
phenomenon can be seen clearly in J1148+5251, where the host galaxy
redshift is 6.419, but the G-P absorption trough only starts at $z =
6.32$ (Walter et al. 2003).

This emission extending below the rest Ly$\alpha$ line implies that
the QSO must be surrounded by a region of ionized gas, presumably
generated by the ionizing radiation from the QSO itself, which allows
photons to 'leak out' to lower redshifts, before encountering the
neutral IGM.  For lower redshift QSOs this is known as the 'proximity
effect', indicating a deficit of Ly$\alpha$ absorption line systems
close to the QSO. In a qualitative sense, we are witnessing the
process of reionization, as the QSO ionizes the IGM in its vicinity.
More quantitatively, the difference in redshift can be used to derive
the size of the cosmic Stromgren sphere generated by the QSO in the
IGM. For J1148+5251 we find a physical radius of R = 4.7 Mpc. 

This is a time bounded Stromgren sphere (or ionization front), from
which one can derive the timescale for the AGN activity from the size
of the sphere, the average IGM baryon density and neutral fraction
(f(HI)), and the QSO ionizing flux (Cen \& Haiman 2000).  For
J1148+5251, Walter et al. (2003) find: $\rm t_{agn} \sim 10^5 ~
R_{Mpc}^3 ~ f(HI) \sim 10^7$ years.\footnote{The ionization front
expands close to the speed of light. However, White et al. (2003) show
that, with a simple change of variables, the relativistic corrections
are implicit in the simple calculation, and the value for $\rm
t_{agn}$ derived from the measured R is the correct proper time
required to generate a sphere of that size.}  A value of $\sim 10^7$
years is considered canonical for AGN activity in luminous QSOs, based
on QSO demographics (Yu \& Tremaine 2002).

Wyithe \& Loeb (2003) have inverted the calculation above to constrain
the IGM neutral fraction.  The simple point is that the timescale
derived above is modulo f(HI).  If f(HI) is as low as $0.001$, as
allowed by the GP measurements, then the AGN lifetime would fall to
$10^4$ years.  Wyithe et al. (2005) have considered the probability of
observing Stromgren spheres of given radii, under the assumption of a
fiducial QSO lifetime between 10$^6$ and 10$^7$ years.  They analyze
the difference between the host galaxy redshift (based on CO or MgII
redshifts, in most cases), and the redshift for the on-set of the GP
absorption trough, for 7 QSOs at $z > 6$.  They conclude that, if the
fiducial lifetime is $> 10^6$ years, the IGM neutral fraction must be:
f(HI)$> 0.1$ to high probability (90th percentile).

A complementary analysis, based on cosmic 'Stromgren surfaces', was
made recently by Mesinger \& Haiman (2004). They compare the redshift
for the onset of GP absorption in the Ly$\alpha$ and Ly$\beta$
lines. Any difference observed is interpreted as the damping wing of
the highly saturated Ly$\alpha$ line arising at the hard edge of the
Stromgren sphere, decreasing the apparent size of sphere as seen in
Ly$\alpha$. From observations of J1030+0524, they also conclude that the
neutral fraction at $z \sim 6.2$ must be: f(HI)$> 0.1$. 

\begin{figure}[htb]
\psfig{figure=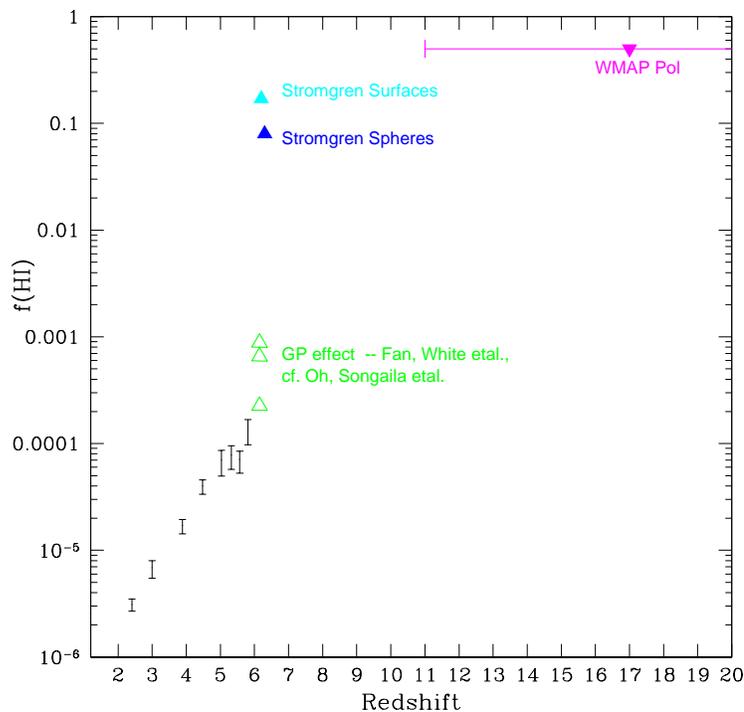,width=4in}
\caption{Current measurements of, or limits to,
the IGM neutral fraction vs. redshift.
Triangles indicate upper or lower limits. 
}
\end{figure}

Figure 5 summarizes the current limits on the IGM neutral fraction at
high $z$. The GP results suggest a significant rise in the neutral
fraction at $z \sim 6$ to 6.5, from f(HI)$\sim 10^{-4}$ to $\ge
10^{-3}$.  The Stromgren spheres and surfaces indicate a very rapid
rise, up to f(HI)$\sim 0.1$ in this redshift range. However, this
rapid rise must be reconciled with the significant ionization fraction
at high redshift derived from the large scale CMB polarization, with
f(HI) $\le 0.5$ at $z \sim 15$. 

It is important to emphasize that all of the measurements in Figure 5
are observationally difficult, and the conclusions are currently a
topic of debate in the literature. I have discussed potential problems
with the GP turn-up and the WMAP turn-down in section 1.  The more
sophisticated analyses of QSO spectra (the Stromgren spheres and
surfaces) are based on only a handful of admittedly noisy spectra,
taken in a particularly problematic wavelength range due to night sky
lines, and Oh \& Furlanetto (2004) have recently challenged both these
conclusions.

A major challenge in the coming years for observational cosmology will
be to "connect-the-dots" between the $z \sim 6$ GP points and the $z
\sim 15$ WMAP point in Figure 5.  However, given the limitations and
difficulties of the current measurements, it is imperative that we
develop an alternative, more direct measure, of the ionization state
of the IGM.  This measure will come in the form of direct imaging of
the IGM through the HI 21cm line of neutral hydrogen.

\section{HI 21cm studies of the neutral IGM during reionization}

The ultimate test of cosmic reionization will come through direct
imaging of the IGM through the HI 21cm line with future large area low
frequency radio telescopes operating in the 100 to 200 MHz range.  The
important point is that, at some time in the history of the universe,
the entire IGM is comprised of neutral hydrogen and helium, thereby
raising the possibility of studying large scale structure (LSS) in
the HI line. Note that this is not LSS in the classic sense of
clusters and superclusters.  The HI 21cm signal from LSS during the
EoR is a combination of density fluctuations, ionization structure,
and excitation (ie. spin) temperature.

\begin{table}[htb]
\caption{EoR HI 21cm Experiments}
\begin{center}
{\small
\begin{tabular}{ccccccccc}
\tableline
\noalign{\smallskip}
experiment & site & $\nu$ range &  FoV & Area & 
B$_{max}$ & cost & date & Goal \\
~ & ~ & MHz & deg & m$^2$ & km & \$ & ~ & ~ \\
\noalign{\smallskip}
\tableline
\noalign{\smallskip}
SKA & ? & 100-200 & 15 & $>$1e6 & 5km=50$\%$ & 1G &  2015? & Imaging \\
LOFAR & NL & 115-240 & 10 & 1e5 & 2km=30$\%$ & 70M & 2007 & PS/CSS \\
MWA/LFD & Oz & 80-300 & 25 & 2e4 & 1.5 & 5M &  2007 & PS/CSS \\
PAST & China &  70-200 & 10 & 8e4 & 10 & ? &  2006 & PS \\
PAPER & USA & 125-175 & 50 & 100 & 0.2 & small & 2006 & $\Delta$T$_{bg}$ \\
MarkIV & Oz &  100-200 & 100 & 1 & 0 & small & 2006 & $\Delta$T$_{bg}$ \\
\tableline
VLA-VHF & USA & 178-204 & 4 & 1.3e4 & 1 & 0.1M & 2005 & CSS (PS) \\
\noalign{\smallskip}
\tableline
\end{tabular}
}
\end{center}
\end{table}

Table 1 lists some current, or planned, experiments to search for the
HI 21cm signal from cosmic reionization. These range from
multi-million dollar programs that will take $>5$ years, to 100
kdollar programs that will be on-the-air within the year. Programs
being pursued include power spectral analyses of the 21cm signal (PS),
searches for cosmic Stromgren spheres associated with the highest $z$
SDSS QSOs (CSS; see section 4.3), large area surveys to look for the
spectral change in the sky brightness temperature as a function of
frequency ($\Delta$T$_{bg}$), and eventual tomographic imaging of the
IGM structure.

Before discussing the HI signatures, I should emphasize that these are
difficult measurements to make for many reasons. First the sky is both
hot and confused at low frequency, with the brightness temperature in
the colder regions of the sky behaving as $\sim 100 ({\nu\over{200 \rm
MHz}})^{-2.6}$ K. The foreground has a large scale component due to
the Galaxy, plus a contribution from many (mostly extragalactic) point
sources.  Moreover, this foreground can become polarized at low
frequency due to differential Faraday rotation through the Galactic
plane (Haverkorn 2004).  Second, the 100 to 200 MHz frequency range is
riddled with terrestrial interference signals, ranging from FM radio
to TV. And third, the ionosphere causes phase fluctuations that
increase as $\nu^{-2}$.

\subsection{Emission}

The simplest experiment entails imaging the full sky to search for a
global step in the brightness temperature at frequencies corresponding
to reionization.  Fig. 6 shows the latest calculation of the expected
HI 21cm signature at low frequencies (Gnedin \& Shaver 2004),
corresponding to a full sky, or very large area measurement.  In this
simulation, at $z > 17$ the signal is zero, since the HI spin
temperature equals the CMB. At $z \sim 13$ resonant scattering of
Ly$\alpha$ photons couples the spin temperature to the kinetic
temperature, which is initially well below the CMB temperature, and the
HI signal is seen in absorption against the CMB. Eventually the IGM
warms up, leading to an emission signature at $z \sim 9$, This
emission signal eventually disappears once the IGM becomes reionized.
However, the full sky measurement is particularly difficult, since the
expected spectral perturbation on the background is at most $\sim 20$
mK.  Gnedin \& Shaver point at that one has to measure changes in the
spectral index of the non-thermal foreground emission to a part in
10$^4$ (Fig. 6).

\begin{figure}[htb]
\psfig{figure=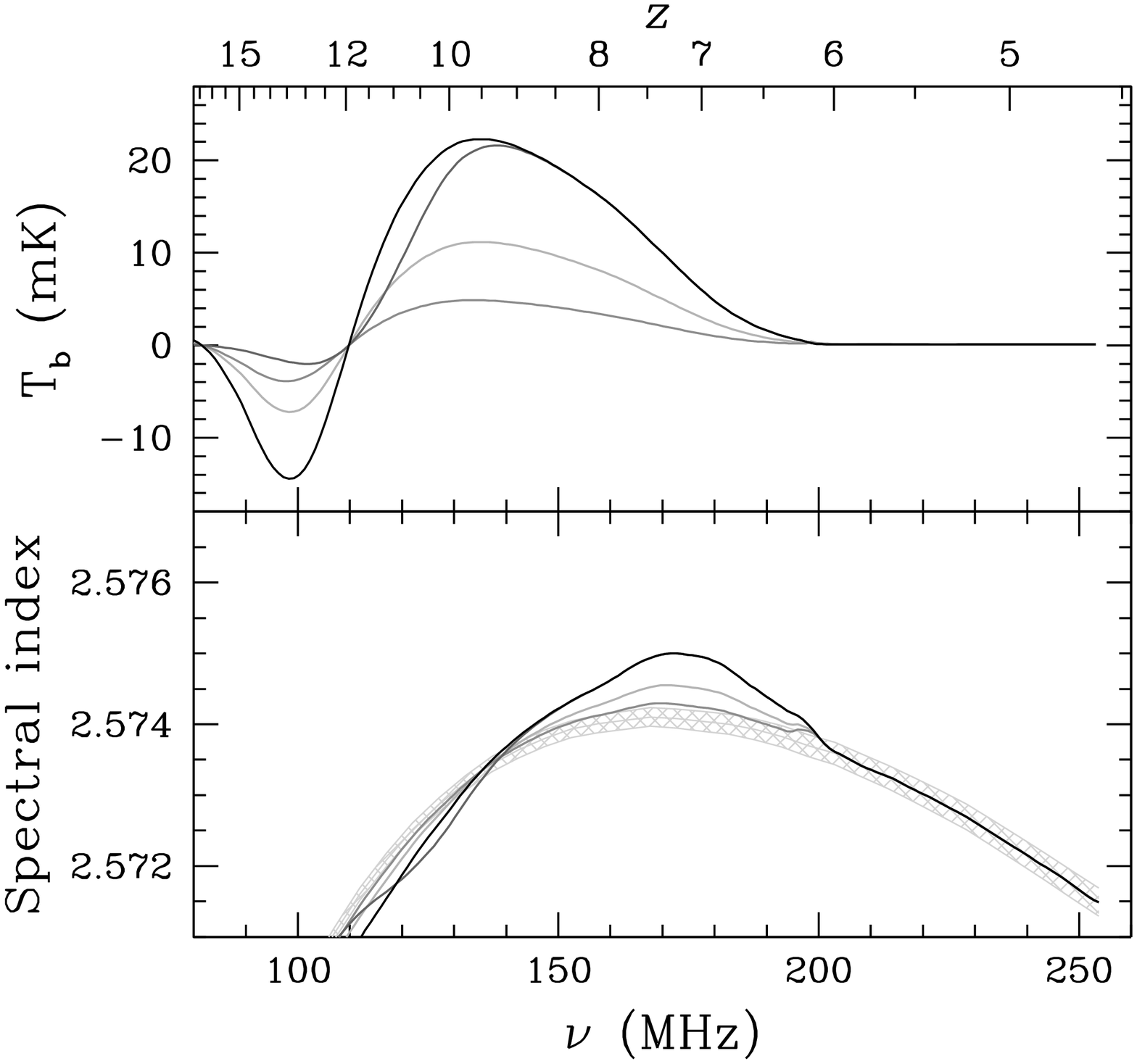,width=2.7in}
\vspace*{-2.7in}
\hskip 2.8in
\psfig{figure=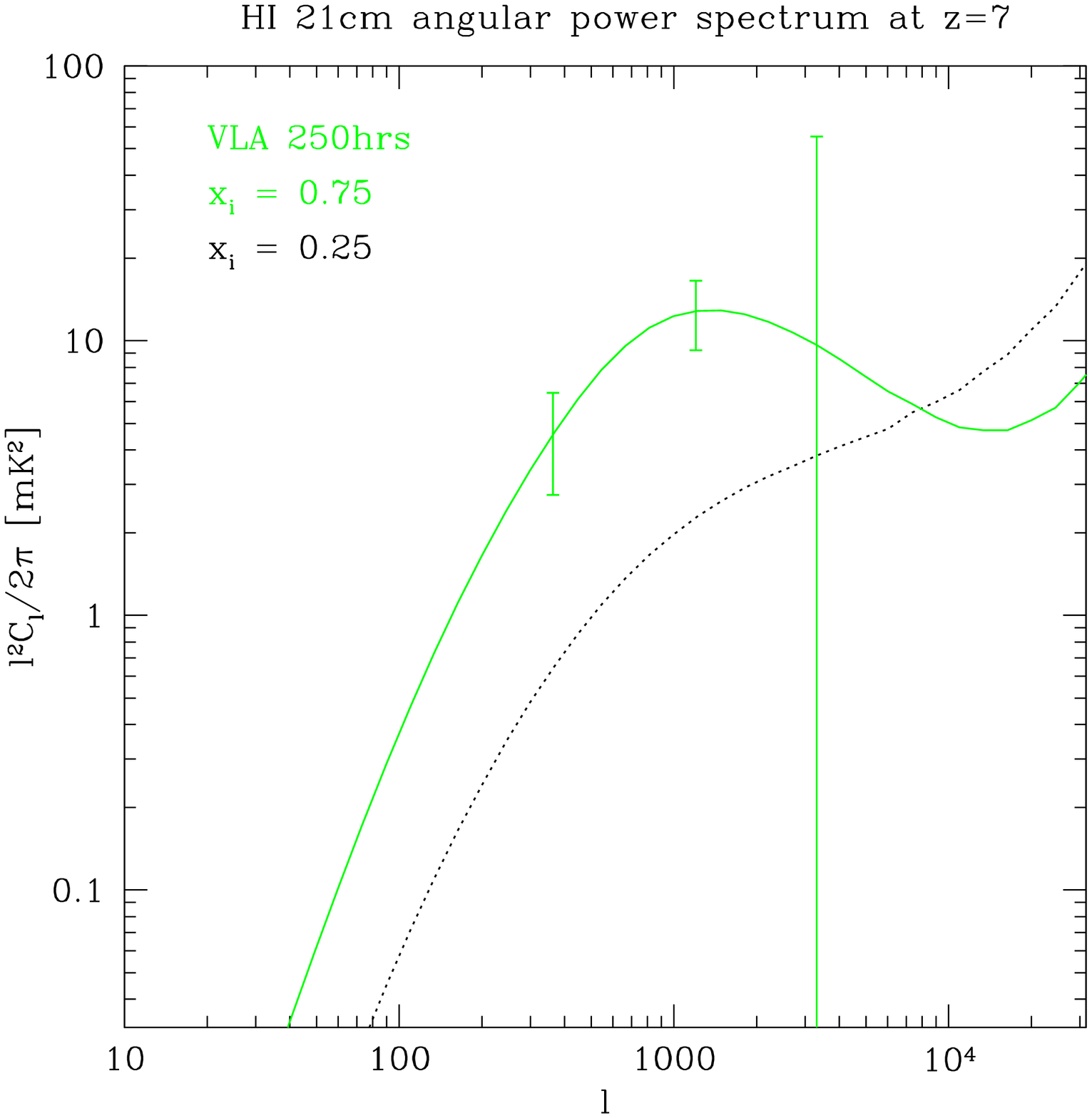,width=2.7in}
\caption{{\bf Left:} The expected HI 21cm global (ie. large area) signal
from cosmic reionization (Gnedin \& Shaver 2004). The darkest line is for
single reionization and the lightest line is for double
reionization. {\bf Right:} Simulated power spectra of HI 21cm emission
during cosmic reionization at $z\sim 6.5$ to 7. The solid (green) line
is for an ionization fraction of 0.75, while the dotted black line is
for 0.25. The error bars indicate the expected sensitivity in a single
1200 km s$^{-1}$ channel for the VLA-VHF system in 250 hours,
including cosmic variance (thanks to L.Greenhill, M.Zaldarriaga). }
\end{figure}

\begin{figure}[htb]
\psfig{figure=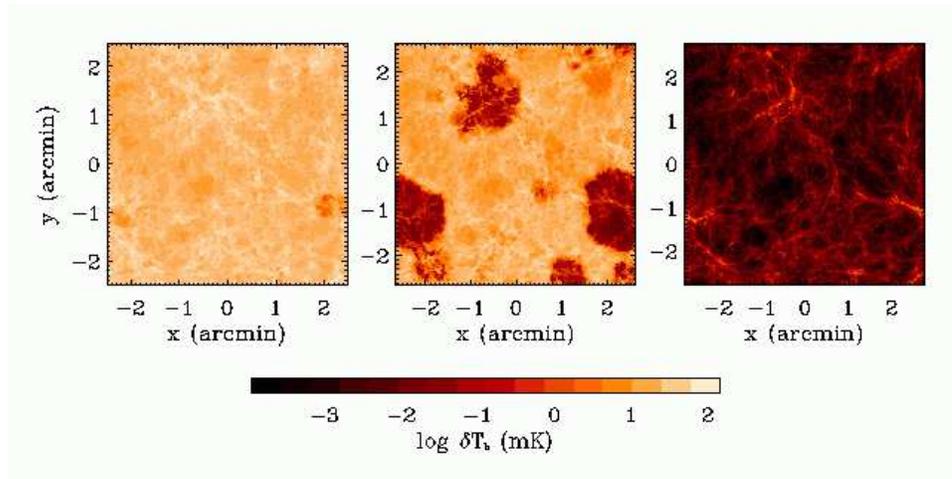,width=5in}
\caption{Simulated images of HI 21cm emission during cosmic reionization 
(Zaldariagga et al. 2004). }
\end{figure}

The ultimate goal of HI 21cm observations is tomographic imaging of
the IGM during reionization (see Furlanetto \& Briggs 2004c
and references therein). Fig 7 shows one example of the type of
signal one might expect. The first galaxies and AGN generate isolated
Stromgren spheres, which eventually grow to overlap, or percolate,
leading to a rapid fall in the neutral fraction, ie. a 'cosmic phase
transition' (Gnedin 2005). The expected signals for the Stromgren
spheres around normal galaxies is about 10mK on scales of 1$'$. The
proposed SKA should have adequate sensitivity to image such
structures, with an expected rms in 100 hours of 4 mK on these scales.
Unfortunately, pathfinder experiments such as LOFAR, PAST, and the MWA
will have insufficient sensitivity to perform such imaging.

However, all hope is not lost. The path finder finder experiments
should have enough sensitivity to measure the power spectrum of the HI
21cm fluctuations.  The direct analogy can be drawn with COBE and
WMAP, in which COBE lacked the sensitivity to image the CMB
fluctuations, but was able to measure the statistical signal via the
power spectrum, while WMAP generated true images of the CMB
fluctuations. Fig. 6b shows the predicted power spectrum in the HI 21cm
fluctuations at $z = 6.5$ to 7, along with the expected errors for
measurements made with the VLA-VHF system (see section 4.3).
An interesting feature of the power spectrum is that the fluctuation 
signal increases as the neutral fraction drops, until some point
when the IGM is almost fully ionized. 

Generating the HI 21cm images and power spectra above requires removal
of the non-thermal foreground to a part in 10$^4$. The hope is that
the foreground can be removed spectrally, since the non-thermal
sources are well known to have  smooth (power-law) spectra over 10's
of MHz (eg. Morales \& Hewitt 2003).

\subsection{Absorption}

Another potential probe of the neutral IGM is through absorption
toward the first radio loud AGN. Fig 8 shows the predicted 21cm
absorption signal toward discrete radio sources at $z = 12$ and $z= 8$
(Carilli et al. 2002). Two signatures are seen: an overall depression
due to the mean neutral IGM of $\sim 0.1\%$ at $z = 12$, and discrete
lines of few km s$^{-1}$ width and few$\%$ optical depth due to
moderate density enhancements, $\delta \sim 10$, expending in the
linear regime. This '21cm forest' corresponds directly to the
Ly$\alpha$ forest after reionization. In essence, after
reionization the Ly$\alpha$ forest clouds have column densities
between 10$^{13}$ and 10$^{15}$ cm$^{-2}$, with a neutral fraction
$\sim 10^{-4}$ to $10^{-5}$. Before reionization this implies
neutral column densities $\sim 10^{19}$ to $10^{21}$ cm$^{-2}$,
adequate for studying the HI 21cm line.  By redshift 7 the HI 21cm
absorption signals have decreased due to the decreased neutral
fraction, and the increased spin temperature.

\begin{figure}
\psfig{figure=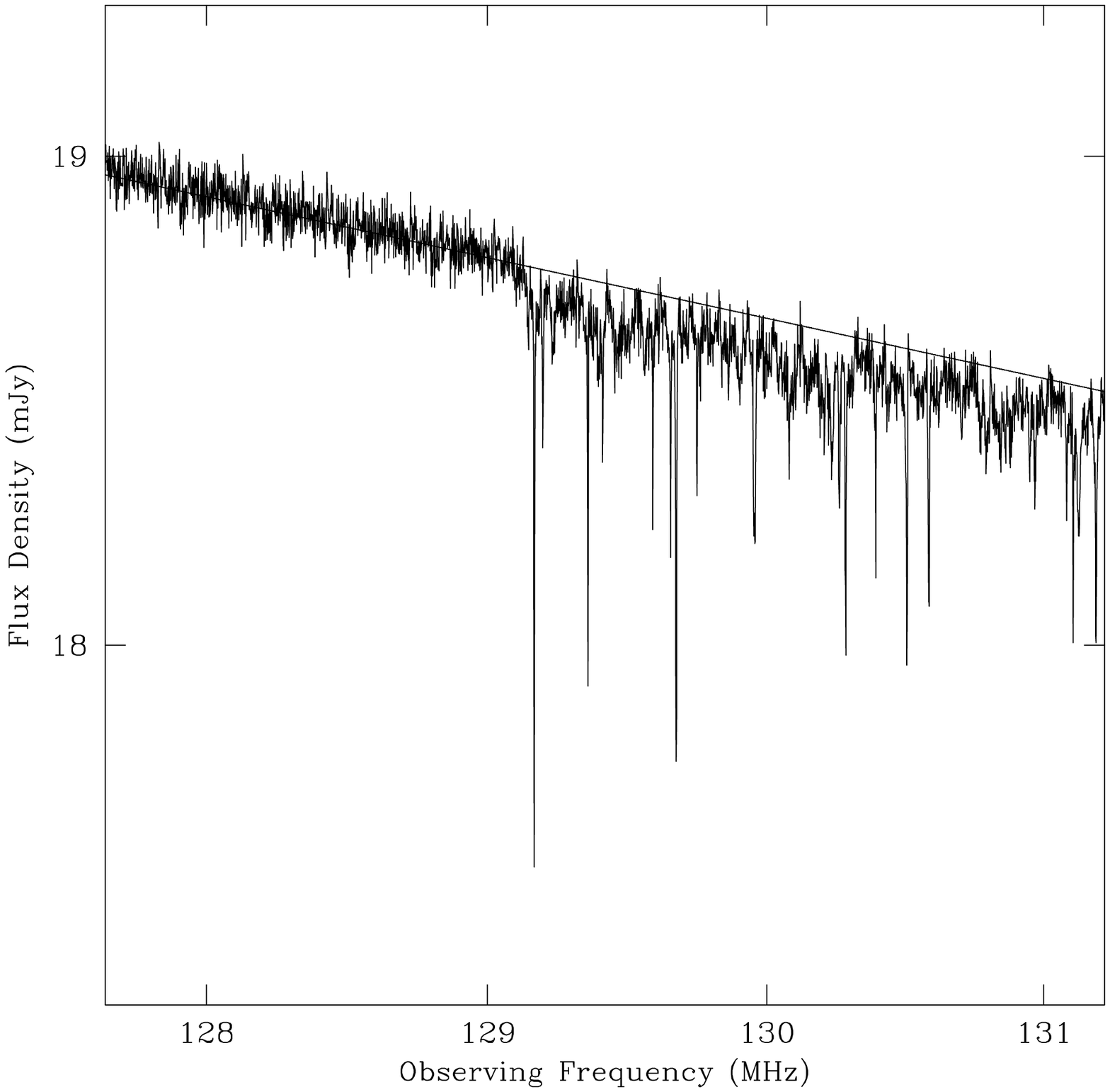,width=2.7in}
\vspace*{-2.7in}
\hskip 2.7in
\psfig{figure=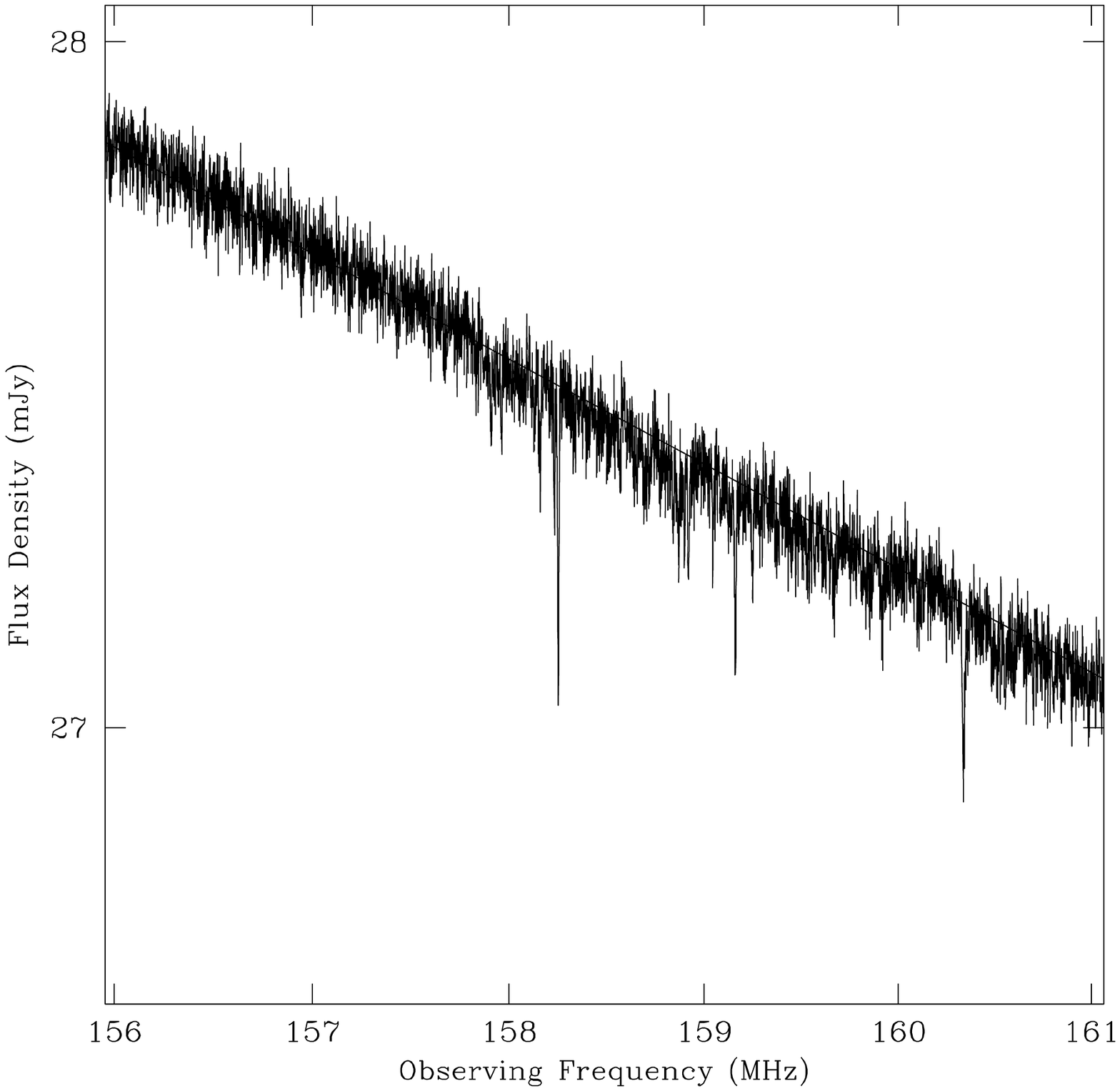,width=2.7in}
\caption{{\bf Left:} the expected HI 21cm absorption spectrum for
a radio loud AGN at $z = 12$, including noise appropriate 
for the SKA in 24 hours. {\bf Right:} The same as 9a, but for $z = 8$
(Carilli et al. 2002).
}
\end{figure}

Furlanetto \& Loeb (2002) have also hypothesized HI 21cm absorption
lines due to the first collapsed objects (mini-halos), and 
proto-disk galaxies. The mini-halos may contribute a similar 
line density as the 21cm forest. The proto-disks will be
rare, but have higher columns, and hence might be able to be
seen against fainter radio sources, such as GRB radio after glows
within the host galaxy. 

Detecting HI 21cm absorption requires a source population at very high
$z$.  A number of authors have considered this question, and conclude
that an adequate number of radio sources should exist within the EoR
to perform such experiments, with a predicted source areal density
between 0.05 and 0.5 radio sources per square degree at $z > 6$ with
S$_{\rm 150MHz} > 6$mJy (Carilli et al. 2002; Haiman et al. 2004;
Rawlings \& Jarvis 2004).  The basic point is that the radio loud AGN
fraction remains a constant 10$\%$ out to the highest redshifts
(Petric et al. 2004). The difficult part will be identifying these
radio sources as very high $z$ objects, although the HI 21cm signal
itself may be useful in this regard (Carilli et al. 2002).

\subsection{On the fast track: VLA imaging of cosmic Stromgren spheres}

I would like to close with a short summary of the VLA-VHF experiment
to study the EoR HI 21cm signal.  This is a joint SAO-NRAO program to
outfit the VLA with prime focus dipole receivers at 178 to 204 MHz for
EoR studies (PI L. Greenhill). Funding for the hardware has been
granted from SAO, and NRAO has agreed to support the first testing of
the system. The project is highly leveraged on existing
infrastructure, and hence can be done both very cheaply ($\sim 100$K
dollars), and very quickly. We will test the first prototypes in
February and March 2005, with the plan of performing the first
experiments with the full array at the end of 2005.  The VLA FoV is
$\sim 12$ deg$^2$, with a resolution of $\sim 7'$ in D array, and an
expected rms per 0.8 MHz channel of 0.1 mJy in 250 hours.

\begin{figure}
\psfig{figure=VLAD-EOR-250.PS,width=2.7in}
\vspace*{-2.9in}
\hskip 2.8in
\psfig{figure=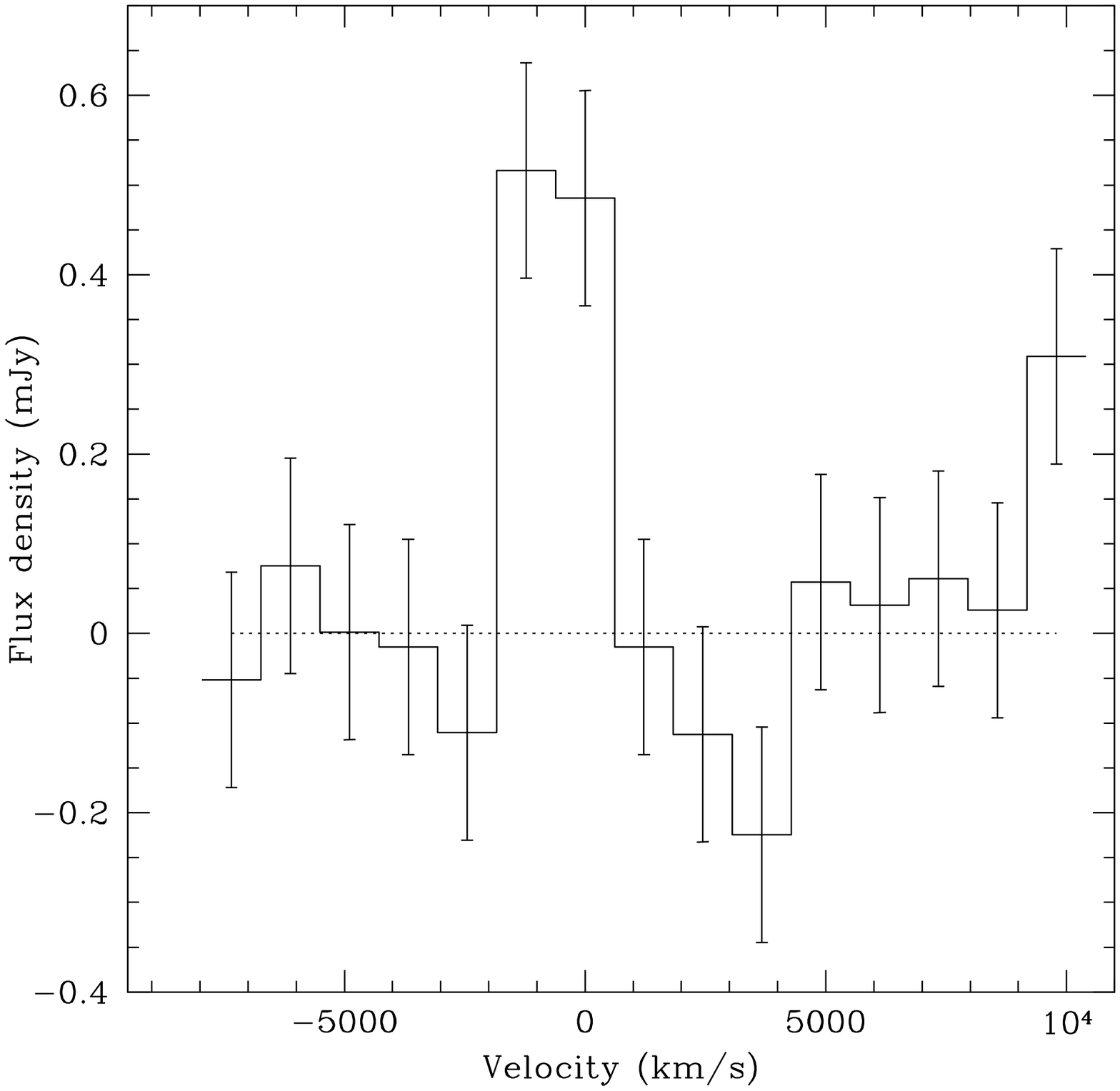,width=2.7in}
\caption{{\bf Left:} the simulated HI 21cm image of the cosmic 
Stromgren sphere around an SDSS $z\sim 6.3$ QSO for f(HI)=1,
assuming 250 hours with the VLA-VHF system. The circle shows
the true location and size of the model. The contour levels
are --0.45, --0.3, --0.15, 0.15, 0.3, 0.45 mJy beam$^{-1}$. 
{\bf Right:} The simulated spectrum for the VLA-VHF observations. 
}
\end{figure}

The SAO/NRAO program has a very specific goal -- to image the cosmic
Stromgren spheres around the highest redshift SDSS QSOs.  Wyithe \&
Loeb (2004b) have predicted the HI 21cm signal from such spheres. They
predict Stromgren spheres of $\sim 15'$ with a brightness temperature
of $\sim 20$f(HI) mK. These spheres will be much larger than those
associated with normal galaxies, and hence easier to detect.  This is
perhaps the easiest EoR experiment possible, since we have evidence
that these spheres exist (section 3), and the redshifts and positions
are known accurately. In piggy-back, we can search for the statistical
signal of the neutral IGM in the power spectrum (Fig. 6b).

Fig. 9 shows simulations of what the VLA-VHF system might observe
toward the $z > 6$ QSOs.  The combination of spatial and spectral
information will be key to verifying the reality of the signal.  We
expect a $\sim 4-5\sigma$ signal in two spectral channels, for
f(HI)$\sim 1$. Conversely, by combining channels, and observing a
number of sources, we hope to set limits on the IGM neutral fraction,
at the level of f(HI)$\ge 0.1$.  While this may be well above the
value expected just based on the GP lower limits, it would represent
the first direct limit on the neutral IGM, and probes to the level
expected based on the Stromgren spheres and surfaces.

\acknowledgements The National Radio Astronomy Observatory is a
facility of the National Science Foundation operated under cooperative
agreement by Associated Universities, Inc.. I thank my collaborators,
F. Bertoldi, K. Menten, F. Walter, P. Cox, A. Beelen, X. Fan,
M. Strauss, L. Greenhill, R. Blundell, M. Zaldarriaga, A. Loeb,
S. Furlanetto, N. Gnedin, F. Owen, A. Omont, P. van den Bout,
P. Solomon, S. Wyithe, and many others, for many things.

\end{document}